\documentclass[a4paper,12pt]{article}
\usepackage{amssymb,amsmath}
\usepackage{bm}%
\usepackage{stmaryrd}
\usepackage[dvipdfmx]{hyperref}
\hypersetup{
	colorlinks=true,
	citecolor=blue,
	linkcolor=blue,
	urlcolor=orange,
}
\usepackage{cite}

\def\nn{\nonumber}
\def\Z2{\mathbb{Z}_2^2}%
\def\g{\mathfrak{g}}
\def\G{\mathfrak{G}}

\begin{document}

\title{Irreducible representations of $\mathbb{Z}_2^2$-graded ${\cal N} =2$ supersymmetry algebra and $\mathbb{Z}_2^2$-graded supermechanics}

\author{N. Aizawa\thanks{Present affiliation: Department of Physics, Osaka Metropolitan University} and  S. Doi
	\\[10pt]
	Department of Physical Science, Osaka Prefecture University, \\
	Nakamozu Campus, Sakai, Osaka 599-8531, Japan}

\maketitle
\thispagestyle{empty}

\vfill
\begin{abstract}
	Irreducible representations (irreps) of $\mathbb{Z}_2^2$-graded supersymmetry algebra of ${\cal N}=2$ are obtained by the method of induced representation and they are used to derive $\mathbb{Z}_2^2$-graded supersymmetric classical actions. 
	The irreps are four dimensional for $ \lambda = 0$ where $ \lambda $ is an eigenvalue  of the Casimir element, and  eight dimensional for $\lambda \neq 0.$ 
	The eight dimensional irreps reduce to four dimensional ones only when $\lambda$ and an eigenvalue of Hamiltonian satisfy a particular relation. The reduced four dimensional irreps are used to define $\mathbb{Z}_2^2$-graded supersymmetry transformations and  two types of classical actions invariant under the transformations are presented. 
	It is shown that one of the Noether charges vanishes if all the variables of specific  $\mathbb{Z}_2^2$-degree are auxiliary.
\end{abstract}

%%%%%%%%%%%%%%%%%%%%%%%%%%%%%%%%%%%%%%%%%%%%%%%%%%%%%%%%%%%%%%%%%%%%%%%%%%%%%%%%%%%%%%
\clearpage
\setcounter{page}{1}

\section{Introduction}

Lie superalgebras are a $\mathbb{Z}_2$-graded extension of Lie algebras. 
It has been known since 1960 that the $\mathbb{Z}_2$-grading can be generalized to arbitrary Abelian groups \cite{Ree,RW1,RW2,sch1}. 
Namely, there exist a wide variety of extensions of Lie superalgebras. 
Among others, $\mathbb{Z}_2^n$-graded Lie superalgebras and symmetries generated by them attract some interest recently ($ \mathbb{Z}_2^n := \mathbb{Z}_2 \times \cdots \times \mathbb{Z}_2$,\; $n$ times). 
This is because such symmetries are found in the known quantum systems \cite{Vasiliev,tol,AKTT1,AKTT2} and many simple systems having such symmetries are constructed \cite{tol2,Bruce,BruDup,AAD,AAd2, AKTcl,AKTqu,DoiAi1,DoiAi2,brusigma,bruSG}. 
It is also pointed out the existence of quantum mechanical observables which detect the particles obeying $\Z2$-graded parastatistics \cite{Topp,Topp2}.
This implies that the higher graded symmetries play a certain role in physics, perhaps, they will provide us a new perspective on symmetries in science.   

However, our understanding of higher graded symmetries is still very limited. 
To have better understanding of $\mathbb{Z}_2^n$-graded supersymmetry ($\mathbb{Z}_2^n$-SUSY), consideration of $\mathbb{Z}_2^n$-graded extensions of supersymmetric \textit{classical} mechanics ($\mathbb{Z}_2^n$-supermechanics) is quite useful as it provides a place where physics, representation theory, calculus on $\mathbb{Z}_2^n$-graded variables and $\mathbb{Z}_2^n$-graded geometry encounter. 
We remark that geometry on $\mathbb{Z}_2^n$-graded manifolds is one of the topics under extensive studies in mathematics \cite{CGP1,CGP2,CGP3,CGP4,CoKPo,Pz2nint,BruIbar,BruPon,BruIbarPonc,BruGraRiemann,BruGrabow,CoKwPon,BruIbarPon2,BruGrabow2} and that the representation theory of higher graded superalgebra also attracts mathematical interest \cite{MohSal,NAJap,IsStvdJ,Meyer,AmaAi,Que}.

$\mathbb{Z}_2^n$-supermechanics has been studied only for $n=2$ and ${\cal N} = 1$ so far \cite{AKTcl,AKTqu,DoiAi2} where $\cal N$ counts the number of supercharges in each subspace of $\Z2$-degree $(1,0)$ and $(0,1)$, see the equation \eqref{GradedVS}. 
In the present work, we start exploring $\Z2$-supermechanics of ${\cal N} = 2.$ 
As is well-known, ${\cal N} = 2 $ theories are most fundamental in the standard supersymmetry and supercharges can be taken as a nilpotent operator in this case. 
The nilpotency is a key connecting supersymmetry and differential geometry. 
Therefore, $\Z2$-supermechanics of ${\cal N} = 2 $ is expected to play the same role in higher graded SUSY and  geometry on $\Z2$-graded manifolds. 

In the following sections, we will proceed to explore $\Z2$-supermechanics of ${\cal N} = 2$ as follows. 
The $\Z2$-SUSY algebra is taken from the authors' previous work \cite{AAD} in which an $ {\cal N}$ extension of $\Z2$-graded supersymmetric \textit{quantum} mechanics is discussed. 
Then what we need is a $\Z2$-SUSY transformation acting on $\Z2$-commutative numbers \cite{RW2} which are functions of time. 
Such transformation should be an irreducible representation (irrep) of the $\Z2$-SUSY algebra, so we start our explore with an abstract representation theory of the $\Z2$-SUSY algebra (see \cite{DoiAi2} for ${\cal N} =1$). 
In \S \ref{SEC:IndRep}, representations of $\Z2$-SUSY algebra induced from irreps of the Cartan subalgebra are presented. 
The induced representations are reducible. 
It then turns out  in \S \ref{SEC:DE} and \S \ref{SEC:DEl} that irreps are either four or eight dimensional depending on the eigenvalues of the Casimir element and Hamiltonian (time translation operator). 
The four dimensional irreps act naturally on complex $\Z2$-commutative numbers, thus we take them as $\Z2$-SUSY transformation. 

Having the $\Z2$-SUSY transformation and dynamical variables, we proceed to construct classical actions invariant under the transformation. 
In \S \ref{SEC:Action}, we present two types of actions. One is the theory of higher order time derivative and the other is free theory consisting of the standard kinetic terms. 
The free theory exhibits two striking features of $\Z2$-supermechanics. 
The first one is a conversion of dynamical and auxiliary variables which is also observed in the standard supermechanics. 
The second one is the fact that if all the variables of  $\Z2$-degree $(0,0)$ or $(1,1)$ is auxiliary, then the conserved Noether charge of degree $(1,1)$ vanishes. 
We comment on some possible future works in \S \ref{SEC:CR}.

Before starting our explore,  we give a definition of $\Z2$-graded Lie superalgebra. 
Let $ \g$ be a vector space and $ a, b \in \Z2.  $ 
Further, let $ a \cdot b$ be the standard inner product of two dimensional vectors. 
Suppose that $\g$ is a direct sum of graded components
\begin{eqnarray}
	 \g = \g_{(0,0)} + \g_{(1,1)} + \g_{(1,0)} + \g_{(0,1)} \label{GradedVS}
\end{eqnarray}
and we denote a homogeneous element of $ \g_a$ by $ X_a.$ 
Then $\g$ is referred to as a $\Z2$-graded Lie superalgebra if it admits the general Lie bracket $ \llbracket X_a, X_b \rrbracket \in \g_{a+b} $ satisfying 
\begin{eqnarray}
	\llbracket X_a, X_b \rrbracket = -(-1)^{a\cdot b} \llbracket X_b, X_a \rrbracket
\end{eqnarray}
and the $\Z2$-graded version of the Jacobi identity which we do not need in the present work. 
One may readily see that the general bracket is realized by commutator (anticommutator) for $ a \cdot b$ even (odd). 
If the general bracket vanishes, i.e., $ \llbracket X_a, X_b \rrbracket = 0$, then we say that $ X_a $ and $X_b$ are $\Z2$-\textit{commutative}.

%%%%%%%%%%%%%%%%%%%%%%%%%%%%%%%%%%%%%%%%%%%%%%%%%%%%%%%%%%%%%%%%%%%%%%%%%%%%%%%%%%%%
%
\setcounter{equation}{0}
\section{Induced representations of $\Z2$-SUSY algebra} \label{SEC:IndRep}

The $\Z2$-SUSY algebra of ${\cal N}=2$, denoted by $ \G $,  has the following elements:
\begin{eqnarray}
	(0,0)\ ; \ H \qquad (1,1)\ ; \ Z \qquad (1,0)\ ; \ Q_{10},\, Q_{10}^{\dagger} \qquad (0,1)\ ; \ Q_{01},\, Q_{01}^{\dagger}
\end{eqnarray}
They satisfy the relations
\begin{align}
	\{ Q_a, Q_a \} &= \{ Q_a^{\dagger}, Q_a^{\dagger} \} = \{Z, Q_a\} = \{Z, Q_a^{\dagger} \} = 0,
	\nonumber \\
	\{ Q_a, Q_a^{\dagger} \} &= H, 
	\quad
	[Q_{01}, Q_{10}^{\dagger}] = [Q_{01}^{\dagger}, Q_{10}] = iZ,
	\nonumber \\
	[Q_a, Q_b] &= [Q_a^{\dagger}, Q_b^{\dagger}] = 0, \quad \forall a, b
	\label{Def:SUSY}
\end{align}
and the Hamiltonian $H$ commute with all other elements. 
It follows from these relations that $Z^2$ is the Casimir element of $\G$.  
Thus, each irrep of $\G $ is labelled by the eigenvalues of $H $ and $Z^2.$ 
In the present work, we consider only the representations on a $\Z2$-graded vector space over $\mathbb{C}. $ 

In order to obtain irreps of $\G$ we consider the representations induced from irreps of the Cartan subalgebra  which is spanned by $H$ and $Z.$ 
It is known that irrep of the Cartan subalgebra is either one dimensional or two dimensional \cite{AmaAi}. 
Denoting the eigenvalue of $H$ by $E$, the one dimensional irrep is given by
\begin{eqnarray}
	H v_1 = E v_1, \qquad Z v_1 = 0
\end{eqnarray}
and the two dimensional irrep is given by
\begin{eqnarray}
	H v_1 = E v_1, \qquad Z v_1 = u_5, \qquad Z u_5 = \lambda v_1, \quad \lambda \neq 0 \in \mathbb{C} \label{Cartan2D}
\end{eqnarray}
where the $\Z2$-degree of $v_1$ is $(0,0) $ and the unusual suffix of $u_5$ is for later convenience. 
One may see from \eqref{Cartan2D} that $\lambda $ is the eigenvalue of $Z^2$. 
Thus the one dimensional irrep corresponds to the zero eigenvalue $(\lambda = 0)$ of $Z^2.$ 

The representations of $\G$ induced from the one and two dimensional irreps are of 16 and 32 dimension, respectively. 
We denote the 16 dimensional representation by $ D(E)$ and 32 dimensional one by $ D(E,\lambda).$  
The basis of $D(E, \lambda)$ is taken as follows. 

\medskip\noindent 
The degree $(0,0)$ vectors:
\begin{alignat}{3}
	v_1&, &  v_2 &= [Q_{10}, Q_{10}^{\dagger}] v_1, & \quad v_3 &= [Q_{01}, Q_{01}^{\dagger}] v_1, 
	\nn \\
	v_4 &= \frac{1}{2} \{ Q_{01}, Q_{10}^{\dagger} \} \{ Q_{01}^{\dagger}, Q_{10} \}v_1,
	& \quad 
	v_5 &= \{ Q_{01}, Q_{10} \} u_5, &
	v_6 &= \{ Q_{01}^{\dagger}, Q_{10}^{\dagger} \} u_5, 
	\nn \\
	v_7 &= \{ Q_{01}, Q_{10}^{\dagger} \} u_5, &
	v_8 &= \{ Q_{01}^{\dagger}, Q_{10} \} u_5.
\end{alignat}
The degree $(1,1)$ vectors:
\begin{alignat}{3}
	u_1 &= \{ Q_{01}, Q_{10} \} v_1, & \qquad 
	u_2 &= \{ Q_{01}^{\dagger}, Q_{10}^{\dagger} \} v_1, & \qquad
	u_3 &= \{ Q_{01}, Q_{10}^{\dagger} \} v_1,
	\nn \\
	u_4 &= \{ Q_{01}^{\dagger}, Q_{10} \} v_1, &
	u_5 &, &
	u_6 &= [Q_{10}, Q_{10}^{\dagger}] u_5, 
	\nn \\
	u_7 &= [Q_{01}, Q_{01}^{\dagger}] u_5, &
	u_8 &= \frac{1}{2} \{ Q_{01}, Q_{10}^{\dagger} \} \{ Q_{01}^{\dagger}, Q_{10} \} u_5.
\end{alignat}
The degree $(1,0)$ vectors:
\begin{alignat}{3}
	\chi_1 &= Q_{10} v_1, & 
	\chi_2 &= Q_{10}^{\dagger} v_1, & 
	\chi_3 &= Q_{01}^{\dagger} \{ Q_{01}, Q_{10}^{\dagger} \} v_1,
	\nn \\
	\chi_4 &= Q_{01} \{ Q_{01}^{\dagger}, Q_{10} \} v_1, & \qquad
	\chi_5 &= Q_{01} u_5, & \qquad 
	\chi_6 &= Q_{01}^{\dagger} u_5, 
	\nn \\
	\chi_7 &= Q_{10} \{ Q_{01}, Q_{10}^{\dagger} \} u_5, & 
	\chi_8 &= Q_{10}^{\dagger} \{ Q_{01}^{\dagger}, Q_{10} \} u_5
\end{alignat}
The degree $(0,1)$ vectors:
\begin{alignat}{3}
	\sigma_1 &= Q_{01} v_1, & 
	\sigma_2 &= Q_{01}^{\dagger} v_1, & 
	\sigma_3 &= Q_{10} \{ Q_{01}, Q_{10}^{\dagger} \}  v_1,
	\nn \\
	\sigma_4 &= Q_{10}^{\dagger} \{ Q_{01}^{\dagger}, Q_{10} \} v_1, & \qquad
	\sigma_5 &= Q_{10} u_5, & \qquad 
	\sigma_6 &= Q_{10}^{\dagger} u_5,
	\nn \\
	\sigma_7 &= Q_{01}^{\dagger} \{ Q_{01}, Q_{10}^{\dagger} \} u_5, &
	\sigma_8 &= Q_{01} \{ Q_{01}^{\dagger}, Q_{10} \} u_5.
\end{alignat}
The basis of $ D(E)$ corresponds to the restriction to the first four members in each graded subspaces, namely, 
$ v_i, u_i, \chi_i $ and $\sigma_i$ with $i = 1, 2, 3, 4.$  
The action of $\G$ on these vectors is summarized in Appendix. 
Due to the centrality of $H$ all the vectors are the eigenvector of $H$ with the eigenvalue $E.$ 

Both $ D(E)$ and $D(E,\lambda)$ are completely reducible and their irreducible decomposition will be presented in the next sections.

%%%%%%%%%%%%%%%%%%%%%%%%%%%%%%%%%%%%%%%%%%%%%%%%%%%%%%%%%%%%%%%%%%%%%%%%%%%%%%%%%%%%
%
\setcounter{equation}{0}

\section{Irreducible decomposition of $D(E)$} \label{SEC:DE}

$D(E)$ is decomposed into a direct sum of four inequivalent irreps:
\begin{eqnarray}
	D(E) = D^{(1)}(E) \oplus D^{(2)}(E) \oplus D^{(3)}(E) \oplus D^{(4)}(E)
\end{eqnarray}
where each $D^{(k)}$ is of four dimension. 
$H$ is diagonal and $Z$ is represented by the zero matrix. 
The basis of $D^{(1)}(E)$ is taken to be 
\begin{align}
	\Psi^{(1)} &= (v^{(1)}, u^{(1)}, \chi^{(1)}, \sigma^{(1)})
	\nn \\
	&=(E^2 v_1-Ev_2-v_4, u_1, 2E\chi_1 - \chi_4, 2E\sigma_1 -\sigma_3)
\end{align}
and the action of $ Q_a, Q_a^{\dagger}$ is given by
\begin{alignat}{2}
	Q_{10} \Psi^{(1)} &= (E\chi^{(1)}, 0, 0, E u^{(1)}), & \qquad
	Q^{\dagger}_{10} \Psi^{(1)} &= (0, \sigma^{(1)}, v^{(1)}, 0),
	\nn \\
	Q_{01} \Psi^{(1)} &= (E\sigma^{(1)}, 0, Eu^{(1)}, 0), & 
	Q^{\dagger}_{01} \Psi^{(1)} &= (0, \chi^{(1)}, 0, v^{(1)}). 
\end{alignat}
$ D^{(2)}(E), D^{(3)}(E) $ and $ D^{(4)}(E) $ are defined on the following basis and actions:
\begin{align}
	\Psi^{(2)} &= (v^{(2)}, u^{(2)}, \chi^{(2)}, \sigma^{(2)})
	\nn \\
	&=(E^2 v_1+Ev_3-v_4, u_2, 2E\chi_2 - \chi_3, 2E\sigma_2 -\sigma_3)
\end{align}
\begin{alignat}{2}
	Q_{10} \Psi^{(2)} &= (0, \sigma^{(2)}, v^{(2)}, 0), & \qquad
	Q^{\dagger}_{10} \Psi^{(2)} &= (E\chi^{(2)}, 0, 0, E u^{(2)}),
	\nn \\
	Q_{01} \Psi^{(2)} &= (0, \chi^{(2)}, 0, v^{(2)}), & 
	Q^{\dagger}_{01} \Psi^{(2)} &= (E\sigma^{(2)}, 0, Eu^{(2)}, 0). 
\end{alignat}
\begin{eqnarray}
	\Psi^{(3)} = (v^{(3)}, u_3, \chi_3, \sigma_3), \quad v^{(3)} =E(v_2-v_3) + v_4
\end{eqnarray}
\begin{alignat}{2}
	Q_{10} \Psi^{(3)} &= (0, \sigma_3, v^{(3)}, 0), & \qquad
	Q_{10}^{\dagger} \Psi^{(3)}  &= (E\chi_3, 0, 0, Eu_3),
	\nn \\
	Q_{01} \Psi^{(3)} &= (E \sigma_3, 0, Eu_3, 0), &
	Q_{01}^{\dagger} \Psi^{(3)}  &= (0, \chi_3, 0, v^{(3)} ).
\end{alignat}
\begin{equation}
	\Psi^{(4)} =(v_4, u_4, \chi_4, \sigma_4)
\end{equation}
\begin{alignat}{2}
	Q_{10} \Psi^{(4)} &= (E\chi_4, 0, 0, E u_4), & \qquad
	Q^{\dagger}_{10} \Psi^{(4)} &= (0, \sigma_4, v_4, 0), 
	\nn \\
	Q_{01} \Psi^{(4)} &= (0, \chi_4, 0, u_4), & 
	Q^{\dagger}_{01} \Psi^{(4)} &= (E\sigma_4, 0, Eu_4, 0).
\end{alignat}

Inequivalence of the irreps are obvious from the four dimensionality of each $ D^{(k)}(E)$. 
More precisely, the representation space of $D^{(k)}(E)$ is a direct sum of one dimensional subspaces and each subspace has a fixed $\Z2$-degree. Recalling that similarity transformation have to be performed inside each subspace, the transformation matrix is a diagonal one which is not able to connect different irreps.  

%%%%%%%%%%%%%%%%%%%%%%%%%%%%%%%%%%%%%%%%%%%%%%%%%%%%%%%%%%%%%%%%%%%%%%%%%%%%%%%%%%%%
%
\setcounter{equation}{0}

\section{Irreducible decomposition of $D(E, \lambda)$} \label{SEC:DEl}

In this representation, $H$ is also diagonal but $Z$ is non-trivial. 
Irreducible decomposition of $D(E,\lambda)$ depends on the value of $\lambda. $ 
If $ \lambda \neq E^2, $ then $D(E,\lambda)$ is decomposed into two inequivalent irreps
\begin{eqnarray}
	D(E,\lambda) = D^{(1)}(E,\lambda) \oplus D^{(1)}(E,\lambda) \oplus D^{(2)}(E,\lambda) \oplus D^{(2)}(E,\lambda)
\end{eqnarray}
where $ \dim D^{(1)}(E,\lambda) = \dim D^{(2)}(E,\lambda) =  8 $ and their multiplicity is two. 
If and only if $\lambda = E^2,$ $ D^{(1)}(E,\lambda) $ and $ D^{(2)}(E,\lambda) $ have a four dimensional invariant subspace. 
Therefore, $ \G$ has two inequivalent  four dimensional irreps for $ \lambda = E^2.$ 

First, we show that $D(E,\lambda)$ is decomposed into a direct sum of four invariant subspaces. 
This is done by giving the basis of each subspaces and action of $\G$ explicitly. 
The basis of the first $D^{(1)}(E,\lambda)$ is taken to be
\begin{align}
	v^{(1)}_1 &= v_6, \qquad 
	v^{(1)}_2 = -\frac{1}{2}(\lambda-2E^2) v_1 + E v_3 - v_4 + \frac{i}{2}(v_7-v_8),
	\nn \\
	u_1^{(1)} &= u_2, \qquad 
	u_2^{(1)} = \frac{i\lambda}{2} (u_3-u_4) - \frac{1}{2}(\lambda-2E^2)u_5 + E u_7 -u_8,
	\nn \\
	\chi^{(1)}_1 &= i\lambda \chi_2 + 2E\chi_6-\chi_8, 
	\qquad
	\chi^{(1)}_2 = 2E \chi_2 -\chi_3-i\chi_6,
	\nn \\
	\sigma^{(1)}_1 &= -i \lambda \sigma_2 + 2E \sigma_6-\sigma_7, 
	\qquad 
	\sigma^{(1)}_2 = 2E \sigma_2 - \sigma_4 + i \sigma_6 
\end{align}
and those of the second $D^{(1)}(E,\lambda)$ is 
\begin{align}
	\tilde{v}^{(1)}_1 &= \frac{\lambda}{2}(\lambda-2E^2) v_1 + \lambda E v_2 + \lambda v_4 + \frac{i}{2} \lambda (v_7-v_8),
	\qquad
	\tilde{v}^{(1)}_2 = (\lambda -E^2) v_5,
	\nn \\
	\tilde{u}^{(1)}_1 &= \frac{i}{2} \lambda (u_3-u_4) + \frac{1}{2} (\lambda -2E^2) u_5 + E u_6 + u_8,
	\qquad \quad  
	\tilde{u}^{(1)}_2 = \lambda (\lambda-E^2) u_1, 
	\nn \\
	\tilde{\chi}^{(1)}_1 &= \lambda (\lambda-E^2) \chi_1 + \lambda E(\chi_4 - i \chi_5) + i\lambda \chi_7,
	\nn \\
	\tilde{\chi}^{(1)}_2 &= i(\lambda-E^2) (2E\chi_1 - \chi_4 - i \chi_5),
	\nn \\
	\tilde{\sigma}^{(1)}_1 &= -E(\lambda-2E^2) \sigma_1 -E^2 (\sigma_3-\sigma_5) -E \sigma_8,
	\nn \\
	\tilde{\sigma}^{(1)}_2 &= -i\lambda E \sigma_1 + i\lambda \sigma_3 + (\lambda -2E^2) \sigma_5 + E \sigma_8. 
\end{align}
The action of $\G$ on these vectors is common for the first and second $D^{(1)}(E,\lambda)$ so we present only  for the first one. 
Let $ \Psi^{(1)} = (v^{(1)}_1, v^{(1)}_2, u^{(1)}_1, u^{(1)}_2, \chi^{(1)}_1, \chi^{(1)}_2, \sigma^{(1)}_1, \sigma^{(1)}_2), $ then 
\begin{align}
	Q_{10} \Psi^{(1)} &= (\chi^{(1)}_1,0, \sigma^{(1)}_2, 0, 0, v^{(1)}_2, u^{(1)}_2, 0),
	\nn \\
	Q_{10}^{\dagger} \Psi^{(1)} &= (0, i\chi^{(1)}_1+E\chi^{(1)}_2, 0, E\sigma^{(1)}_1+i\lambda \sigma^{(1)}_2, E v^{(1)}_1, -iv^{(1)}_1, -i\lambda u^{(1)}_1, Eu^{(1)}_1),
    \nn \\
	Q_{01} \Psi^{(1)} &= (\sigma^{(1)}_1, 0, \chi^{(1)}_2, 0, u^{(1)}_2, 0, 0, v^{(1)}_2),
    \nn \\
	Q_{01}^{\dagger} \Psi^{(1)} &= (0, -i\sigma^{(1)}_1+E\sigma^{(1)}_2, 0, E\chi^{(1)}_1-i\lambda \chi^{(1)}_2, i\lambda u^{(1)}_1, E u^{(1)}_1, E v^{(1)}_1, i v^{(1)}_1),
    \nn \\
	Z\Psi^{(1)} &= (\lambda u^{(1)}_1,  u^{(1)}_2,  v^{(1)}_1, \lambda v^{(1)}_2, -\lambda \sigma^{(1)}_2, -\sigma^{(1)}_1,  -\lambda \chi^{(1)}_2, -\chi^{(1)}_1).
\end{align}

For a basis of $ D^{(2)}(E,\lambda),$ one may take the following vectors:
\begin{alignat}{2}
	v^{(2)}_1 &= -\frac{\lambda}{2}v_1 + Ev_2 + v_4 - \frac{i}{\lambda}(\lambda-2E^2) v_7 - \frac{i}{2} v_8,
	& \qquad v^{(2)}_2 &= i\lambda v_3 - Ev_7,
	\nn \\
	u^{(2)}_1 &= \frac{\lambda}{2} (iu_3+iu_4+u_5) - E(u_6-u_7) - u_8, 
	&
	u^{(2)}_2 &= iEu_3 + u_7,
	\nn \\
	\chi^{(2)}_1 &= E \chi_1 - \chi_4 + i \chi_5 - \frac{i}{\lambda} E \chi_7, 
	&
	\chi^{(2)}_2 &= \chi_3 - i \chi_6, 
	\nn \\
	\sigma^{(2)}_1 &= -\lambda \sigma_1 + E(\sigma_3+i\sigma_5) - i\sigma_8,
	&
	\sigma^{(2)}_2 &= i\lambda \sigma_2 - \sigma_7
\end{alignat}
and
\begin{align}
	\tilde{v}^{(2)}_1 &= -\frac{i\lambda}{2} E v_1 + i\lambda v_2 + iE v_4 - \frac{E}{2} v_7 + \Big( 1-\frac{E}{2} + \frac{E^2}{\lambda} - \frac{E^3}{\lambda} \Big) v_8,
	\nn \\
	\tilde{v}^{(2)}_2 &= \frac{\lambda^2}{2} v_1 - \lambda v_4 - \frac{i}{2} v_7 -i \Big( \frac{\lambda}{2} - E + E^2 \Big) v_8,
	\nn \\
	\tilde{u}^{(2)}_1 &= -\lambda E u_4 - i\lambda u_6,
	\nn \\
	\tilde{u}^{(2)}_2 &= -\frac{\lambda}{2}(u_3+u_4 + iu_5) + i u_8,
	\nn \\
	\tilde{\chi}^{(2)}_1 &= i \lambda \chi_1 - iE \chi_4 - E\chi_5 + \chi_7,
	\nn \\
	\tilde{\chi}^{(2)}_2 &= i \lambda \chi_2 + \chi_8, 
	\nn \\
	\tilde{\sigma}^{(2)}_1 &= -i\lambda E \sigma_1 + i\lambda \sigma_3 - \lambda \sigma_5 + E \sigma_8,
	\nn \\
	\tilde{\sigma}^{(2)}_2 &= -\lambda \sigma_4 -i\lambda \sigma_6.
\end{align}
The action of $\G$ on these vectors is common for the first and second sets
so we present only the action on the first set. 
Let $ \Psi^{(2)} = (v^{(2)}_1, v^{(2)}_2, u^{(2)}_1, u^{(2)}_2, \chi^{(2)}_1, \chi^{(2)}_2, \sigma^{(2)}_1, \sigma^{(2)}_2), $ then 
\begin{align}
	Q_{10} \Psi^{(2)} &= ( -E \chi^{(2)}_1, -i\lambda \chi^{(2)}_1, 0, i\sigma^{(2)}_1, 0, v^{(2)}_1 + \frac{i}{\lambda} E v^{(2)}_2, 0, u^{(2)}_1 ),
	\nn \\
	Q_{10}^{\dagger} \Psi^{(2)} &= ( 0, -i\lambda \chi^{(2)}_2, E\sigma^{(2)}_2, \sigma^{(2)}_2, -v^{(2)}_1, 0, iu^{(2)}_1 -iE u^{(2)}_1, 0 ),
	\nn \\
	Q_{01} \Psi^{(2)} &= ( \sigma^{(2)}_1, 0, -i\lambda \chi^{(2)}_1, 0, 0, -i u^{(2)}_2, 0, v^{(2)}_2 ), 
	\nn \\
	Q_{01}^{\dagger} \Psi^{(2)} &= ( \frac{i}{\lambda}(\lambda-E^2) \sigma^{(2)}_2, E \sigma^{(2)}_2, i\lambda \chi^{(2)}_2, iE \chi^{(2)}_2, \frac{i}{\lambda} E u^{(2)}_1-iu^{(2)}_2, 0, E v^{(2)}_1-\frac{i}{\lambda} (\lambda-E^2) v^{(2)}_2, 0  ),
	\nn \\
	Z\Psi^{(2)} &= ( -u^{(2)}_1+E u^{(2)}_2, i\lambda u^{(2)}_2, -\lambda v^{(2)}_1-iE v^{(2)}_2, -i v^{(2)}_2, i\sigma^{(2)}_1, \sigma^{(2)}_2, -i\lambda \chi^{(2)}_1, \lambda \chi^{(2)}_2  ).
\end{align}
One may observe by direct computation that there is no similarity transformation connecting $ D^{(1)}(E,\lambda) $ and $ D^{(2)}(E,\lambda)$. Therefore, the representations $ D^{(1)}(E,\lambda) $ and $ D^{(2)}(E,\lambda)$ are inequivalent.  

Next, we show that $ D^{(1)}(E,\lambda) $ and $ D^{(2)}(E,\lambda)$ are irreducible if $ \lambda \neq E^2.$ 
This is done by showing that $D^{(k)}(E,\lambda)$ has invariant subspace if and only if $ \lambda = E^2. $ 
The invariant subspace is of four dimension so that  $\G$ has the four dimensional irreps for this particular value of $\lambda. $ 
Recall that the each graded component of $D^{(k)}(E,\lambda)$ is two dimensional and an invariant subspace (if any) is also $\Z2$-graded. 
This implies that each graded component of an invariant subspace in $D^{(k)}(E,\lambda)$ is at most one dimensional. 
Suppose that $ w^{(1)} = c_1 v^{(1)}_1 + c_2 v^{(1)}_2 $ is a vector of $\Z2$-degree $(0,0)$ in the invariant subspace in $D^{(1)}(E,\lambda)$.  
Then, one may produce four vectors of degree $(1,0)$ by the action of $\G$:
\begin{alignat}{2}
	Q_{10} w^{(1)}  &= c_1 \chi^{(1)}_1, 
	& \qquad
	Q_{10}^{\dagger} w^{(1)} &= c_2 (i\chi^{(1)}_1 + E\chi^{(1)}_2), 
	\nn \\
	Q_{01} Z w^{(1)} &= c_1 \lambda \chi^{(1)}_2, 
	& \qquad
	Q_{01}^{\dagger} Z w^{(1)} &= c_2 ( E \chi^{(1)}_1 - i\lambda \chi^{(2)}_2).
\end{alignat}
These vectors must be proportional to one another and this requires that $ c_1 = 0, \lambda = E^2. $ 
It follows that the four dimensional space spanned by the vectors 
\begin{eqnarray}
	  v = v^{(1)}_2, \qquad u = u^{(1)}_2, \qquad \chi = i\chi^{(1)}_1 + E \chi^{(1)}_2, \qquad 
	  \sigma = -i \sigma^{(1)}_1 + E \sigma^{(1)}_2 
	  \label{base4D1}
\end{eqnarray}
is invariant under the action of $\G$: 
\begin{alignat}{2}
	Q_{10} \Phi^{(1)} &= (0, 0, Ev, -iu),  & \qquad 
	Q_{10}^{\dagger} \Phi^{(1)} &= (\chi, iE \sigma, 0, 0),
	\nn \\
	Q_{01} \Phi^{(1)} &= (0, 0, iu, Ev), & 
	Q_{01}^{\dagger} \Phi^{(1)} &= (\sigma, -iE \chi, 0, 0),
	\nn \\
	Z \Phi^{(1)} &= (u, E^2 v, -iE \sigma, i E\chi)
\end{alignat}
where $ \Phi^{(1)} = (v, u, \chi, \sigma)$. 
Therefore, we have shown that $ D^{(1)}(E,E^2)$ has a four dimensional invariant subspace iff $\lambda = E^2.$

  One may repeat the same argument for $ D^{(2)}(E,\lambda)$ starting with the degree $(0,0)$ vector 
$ w^{(2)} = c_1 v^{(2)}_1 + c_2 v^{(2)}_2. $ 
The degree $(1,0)$ vectors
\begin{alignat}{2}
	Q_{10} w^{(2)} &= -(E c_1  + i\lambda c_2) \chi^{(2)}_1, 
	& \qquad
	Q_{10}^{\dagger} &= -i\lambda c_2 \chi^{(2)}_2,
	\nn \\
	Q_{01}Z w^{(2)} &= i\lambda c_1 \chi^{(2)}_1, 
	& 
    Q_{01}^{\dagger}Z w^{(2)} &= -(i(\lambda-E^2) c_1 + \lambda E c_2) \chi^{(2)}_2 
\end{alignat} 
must be proportional to one another. 
This requires that one of $c_1 $ and $c_2$ must vanish. 
$ c_1 = 0 $ leads to $ c_2 = 0 $ which means that $ w^{(2)} = 0. $ 
Thus $ c_2 = 0 $ the only possibility and we need further $ \lambda = E^2 $ for this case. 
Then, the four  vectors
\begin{eqnarray}
	\bar{v}  = v^{(2)}_1, \qquad \bar{u} = E^{-1} u^{(2)}_1 - u^{(2)}_2, \qquad 
	\bar{\chi} = \chi^{(2)}_1, \qquad \bar{\sigma} = \sigma^{(2)}_1  \label{bases4D2}
\end{eqnarray}
span an invariant subspace in $ D^{(2)}(E,E^2)$: 
\begin{alignat}{2} 
	Q_{10} \Phi^{(2)} &= (-E \bar{\chi}, -i\bar{\sigma}, 0, 0 ), & \qquad
	Q_{10}^{\dagger} \Phi^{(2)} &= (0, 0, -\bar{v}, iE \bar{u}),
	\nn \\
	Q_{01} \Phi^{(2)} &= (\bar{\sigma}, -iE \bar{\chi}, 0, 0), 
	&
	Q_{01}^{\dagger} \Phi^{(2)} &= (0, 0, i \bar{u}, E \bar{v}),
	\nn \\
	Z \Phi^{(2)} &= (-E \bar{u}, -E \bar{v}, i \bar{\sigma}, -iE^2 \bar{\chi})
\end{alignat}
where $  \Phi^{(2)} = (\bar{v}, \bar{u}, \bar{\chi}, \bar{\sigma}). $ 

 We have shown that $\G$ has four dimensional irreps only when $ \lambda = E^2,$ i.e., when the eigenvalues of $Z$ and $ H^2 $ coincide. 
This relation gives a physical meaning to the parameter $\lambda.$ 
In fact, this also holds true for all the known examples of the $\Z2$-graded version of quantum and classical mechanics \cite{BruDup,AAD,AKTcl,AKTqu,AmaAi}.  
All those works, except \cite{AAD}, discusses the $\Z2$-graded version of  ${\cal N} = 1$ supersymmetry. 
In the next section,  we consider $\Z2$-graded supermechanics of $ {\cal N} = 2 $ under this constraint. 

%%%%%%%%%%%%%%%%%%%%%%%%%%%%%%%%%%%%%%%%%%%%%%%%%%%%%%%%%%%%%%%%%%%%%%%%%%%%%%%%%%%%
%
\setcounter{equation}{0}

\section{Invariant actions} \label{SEC:Action}

In this section, we discuss some possible classical actions which are invariant under the $ \Z2$-SUSY transformation defined by the four dimensional irreps introduced in \S \ref{SEC:DEl}. 
First, we scale the basis \eqref{base4D1} and \eqref{bases4D2} and the scaled basis are also denoted by the same notations:
\begin{equation}
	(v, u, \chi, \sigma)  := (v, E^{-1} u, -\chi, -i\sigma),
	\qquad
	(\bar{v}, \bar{u}, \bar{\chi}, \bar{\sigma}) := (E^{-1} \bar{v}, -E^{-1} \bar{u}, -\bar{\chi}, -iE^{-1} \bar{\sigma}).
\end{equation}
Then, the action of $\G$ on the new basis is given by
\begin{alignat}{2}
	Q_{10} \Phi^{(1)} &= (0, 0, -Ev, -Eu),  & \qquad 
    Q_{10}^{\dagger} \Phi^{(1)} &= (-\chi, -\sigma, 0, 0),
    \nn \\
    Q_{01} \Phi^{(1)} &= (0, 0, -iEu, -iEv), & 
    Q_{01}^{\dagger} \Phi^{(1)} &= (i\sigma, i\chi, 0, 0),
    \nn \\
    Z \Phi^{(1)} &= (Eu, E v, -E \sigma, -E\chi) \label{Action4D1}
\end{alignat}
and
\begin{alignat}{2} 
	Q_{10} \Phi^{(2)} &= (\bar{\chi}, -\bar{\sigma}, 0, 0 ), & \qquad
	Q_{10}^{\dagger} \Phi^{(2)} &= (0, 0, E\bar{v}, -E \bar{u}),
	\nn \\
	Q_{01} \Phi^{(2)} &= (i\bar{\sigma}, -i \bar{\chi}, 0, 0), 
	&
	Q_{01}^{\dagger} \Phi^{(2)} &= (0, 0, iE \bar{u}, -iE \bar{v}),
	\nn \\
	Z \Phi^{(2)} &= (E \bar{u}, E \bar{v}, E \bar{\sigma}, E \bar{\chi}). \label{Action4D2}
\end{alignat}

The basis is a function of $E$ which is interpreted physically as energy. 
To consider classical mechanics, the basis should be a function of time $t.$ 
We thus make a Fourier transform of the basis
\begin{eqnarray}
	(v, u, \chi, \sigma) \to (x(t), z(t), \psi(t), \xi(t)), 
	\quad
	(\bar{v}, \bar{u}, \bar{\chi}, \bar{\sigma}) \to (\bar{x}(t), \bar{z}(t), \bar{\psi}(t), \bar{\xi}(t)) 
	\label{FTvariables}
\end{eqnarray}
where
\begin{equation}
   x(t) = \frac{1}{\sqrt{2}} \int v(E) e^{-iEt} dE, \quad \text{etc.}
\end{equation}
By this transformation, $H$ acts as $ i\partial_t$ as expected. 
The transformed variables \eqref{FTvariables} are treated as $\Z2$-commutative functions of $t.$ 

The $\Z2$-SUSY transformation of the variables in \eqref{FTvariables} is defined by \eqref{Action4D1}, \eqref{Action4D2} and $\Z2$-commutative parameters $ \epsilon, \bar{\epsilon} $ (they are also $\Z2$-commutative with the variables in \eqref{FTvariables}):
\begin{eqnarray}
  \delta_{10} = - \epsilon_{10} Q_{10} - \bar{\epsilon}_{10} Q^{\dagger}_{10},
\quad
\delta_{01} = - \epsilon_{01} Q_{01} - \bar{\epsilon}_{01} Q^{\dagger}_{01},
\quad
\delta_{11} = i \epsilon_{11} Z
\label{Def:variation}
\end{eqnarray}
where $ \bar{\epsilon}_{10}$ is the conjugate of $ \epsilon_{10} $ and $\bar{\epsilon}_{11} = \epsilon_{11}.$
More explicitly,
\begin{align}
	\delta_{10} (x, z, \psi, \xi) &= (\bar{\epsilon}_{10} \psi, \bar{\epsilon}_{10} \xi, i\epsilon_{10} \dot{x}, i\epsilon_{10} \dot{z}),
	\nn\\
	\delta_{01} (x, z, \psi, \xi) &= (-i\bar{\epsilon}_{01} \xi, -i\bar{\epsilon}_{01} \psi, -\epsilon_{01} \dot{z}, -\epsilon_{01} \dot{x}  ),
	\nn \\
	\delta_{11} (x, z, \psi, \xi) &= (-\epsilon_{11} \dot{z},-\epsilon_{11} \dot{x}, \epsilon_{11} \dot{\xi}, \epsilon_{11} \dot{\psi} ) \label{SUSYTF1}
\end{align}
and
\begin{align}
	\delta_{10} (\bar{x}, \bar{z}, \bar{\psi}, \bar{\xi}) &= (-\epsilon_{10} \bar{\psi}, \epsilon_{10} \bar{\xi}, -i\bar{\epsilon}_{10} \dot{\bar{x}}, i\bar{\epsilon}_{10} \dot{\bar{z}}),
	\nn \\
	\delta_{01} (\bar{x}, \bar{z}, \bar{\psi}, \bar{\xi}) &= (-i\epsilon_{01} \bar{\xi}, i\epsilon_{01} \bar{\psi}, \bar{\epsilon}_{01} \dot{\bar{z}}, -\bar{\epsilon}_{01} \dot{\bar{x}}),
	\nn \\
	\delta_{11} (\bar{x}, \bar{z}, \bar{\psi}, \bar{\xi}) &= (-\epsilon_{11} \dot{\bar{z}}, -\epsilon_{11} \dot{\bar{x}}, -\epsilon_{11} \dot{\bar{\xi}}, -\epsilon_{11} \dot{\bar{\psi}}).  \label{SUSYTF2}
\end{align}
One may see that the variables with the bar ``$\bar{\, \ \,}$" are the conjugate of the ones without it.

Having the $\Z2$-SUSY transformation, we present two types of classical actions invariant under \eqref{SUSYTF1} and \eqref{SUSYTF2}. 
The first type is quite general. 
Let $ g(x,z,\psi,\xi,\bar{x},\bar{z},\bar{\psi}, \bar{\xi})$ be a differentiable function of degree $(1,1).$ 
Then, the following is $\Z2$-graded supersymmetric:
\begin{eqnarray}
	 S = \int L\, dt, \qquad
	 L= Z Q_{10}^{\dagger} Q_{10} Q_{01}^{\dagger} Q_{01}  g. \label{Action1}
\end{eqnarray}
The invariance of $S$ readily follows from the relations \eqref{Def:SUSY} and $ Z^2 = H^2 =-\partial_t^2. $ 
For instance, 
\begin{eqnarray}
	 Q_{10} L = -H Z Q_{01}^{\dagger} Q_{01}  g = -i\frac{d}{dt}ZQ_{01}^{\dagger} Q_{01}  g.
\end{eqnarray}
The Lagrangian \eqref{Action1} generally provides higher derivative theories. 
As an illustration, we make a simple choice of the function $g = \mu |x|^2$ where $\mu$ is a constant of degree $(1,1). $ 
Then, \eqref{Action1} yields
\begin{eqnarray}
	L = 2i\mu(\dot{x} \ddot{\bar{z}} - \ddot{\bar{x}} \dot{z} +i \dot{\bar{\psi}} \dot{\xi} -i \dot{\psi} \dot{\bar{\xi}}).
\end{eqnarray}

One may also find more natural models which are our second type. 
It is immediate to verify that the following $L_0$ is $\Z2$-supersymmetric:
\begin{eqnarray}
	L_0 = \dot{\bar{x}} \dot{x} + \dot{\bar{z}} \dot{z} -i (\bar{\psi} \dot{\psi} + \bar{\xi} \dot{\xi} ). 
	\label{FreeL1}
\end{eqnarray}
This is a free theory, however, one may learn some features of $\Z2$-graded supermechanics from this simple model. 
Here we address two of them. First one is a conversion of dynamical and auxiliary variables, the second one is vanishing of Noether charge. 
The first one is also observed in the standard ($\mathbb{Z}_2$-graded) supermechanics. 
One may exchange freely dynamical and auxiliary variables of degree $(0,0)$ and $(1,1)$ without destroying the nature of the representation basis.  
For instance, one may introduce a new degree $(1,1)$ variables
\begin{eqnarray}
	F := \dot{z}, \qquad \overline{F} := \dot{\bar{z}}
\end{eqnarray}
then the Lagrangian $L_0$ becomes
\begin{equation}
	L_1 = \dot{\bar{x}} \dot{x} + |F|^2 -i (\bar{\psi} \dot{\psi} + \bar{\xi} \dot{\xi} )
\end{equation}
which indicates that the variables $F, \overline{F}$ are not dynamical. 
The $\Z2$-SUSY transformation can be written in terms of $ (x,F,\psi,\xi)$ and their conjugate as immediately seen from \eqref{SUSYTF1} and \eqref{SUSYTF2}. 
This new basis is connected to the original one $ (x, z, \psi, \xi)$  by the dressing transformation discussed in \cite{PT,KRT}. 

One may further convert one or all the degree $(0,0)$ variables to auxiliary. 
The following change of variables produces one physical and one auxiliary bosons:
\begin{eqnarray}
	y := \frac{1}{2} (x+\bar{x}), \qquad A:= \frac{i}{2} (\dot{x} - \dot{\bar{x}})
\end{eqnarray}
Then the Lagrangian  $L_1$  is converted into 
\begin{equation}
	L_2 = \dot{y}^2 + A^2 + |F|^2 -i (\bar{\psi} \dot{\psi} + \bar{\xi} \dot{\xi} )
\end{equation}
and the $\Z2$-SUSY transformation can be written in terms of $(y, A, F, \overline{F}, \psi, \bar{\psi}, \xi, \bar{\psi}) $ only. 
Of course, one may keep the degree $(1,1)$ variables dynamical
\begin{eqnarray}
	L_3 = \dot{y}^2 + A^2 + \dot{\bar{z}} \dot{z} -i (\bar{\psi} \dot{\psi} + \bar{\xi} \dot{\xi} )
\end{eqnarray}
or all the degree $(0,0)$ variables is auxiliary
\begin{equation}
	L_4 = |a|^2 + \dot{\bar{z}} \dot{z} -i (\bar{\psi} \dot{\psi} + \bar{\xi} \dot{\xi} )
\end{equation}
where we introduced the new variables $ a = \dot{x}, \bar{a} = \dot{\bar{x}}. $ 

The second feature is on the degree $(1,1)$ Noether charge. 
It vanishes if  all the degree $(0,0)$ or $(1,1)$ variables are auxiliary. 
We present the Noether charges (up to overall constant multiple) for the above mentioned Lagrangians $ L_i, \ i = 0, 1, \dots, 4 $ except the charge corresponding to the time translation (Hamiltonian). 
We denote the Noether charges by the same notations as the elements of $\G $ which does not cause any confusion.  
The Lagrangian $ L_0 $ has no auxiliary variables and the Noether charges are given by
\begin{alignat}{3}
    Q_{10} &= \dot{x} \bar{\psi} + \dot{z} \bar{\xi}, & \qquad
    Q_{10}^{\dagger} &= \dot{\bar{x}} \psi - \dot{\bar{z}} \xi,
    \nn \\
    Q_{01} &= \dot{x} \bar{\xi} + \dot{z} \bar{\psi}, &
    Q_{01}^{\dagger} &= \dot{\bar{x}}  \xi - \dot{\bar{z}} \psi, & \qquad
    Z &= \dot{\bar{x}} \dot{z} + \dot{x} \dot{\bar{z}}
\end{alignat}
which shows all the charges are non-vanishing. 
While, all the degree $(1,1)$ variables are auxiliary in $L_1$ and the Noether charges are given by
\begin{equation}
	Q_{10} = \dot{x} \bar{\psi},  \qquad
    Q_{10}^{\dagger} = \dot{\bar{x}} \psi, 
    \qquad 
    Q_{01} = \dot{x} \bar{\xi}, \qquad
    Q_{01}^{\dagger} = \dot{\bar{x}} \xi, 
    \qquad Z = 0.
\end{equation}
The Lagrangian $ L_2$ has one more auxiliary variable and the conserved charges are 
\begin{equation}
	Q_{10} = \dot{y} \bar{\psi}, \qquad 
	Q_{10}^{\dagger} = \dot{y} \psi, 
	\qquad
	Q_{01} = \dot{y} \bar{\xi}, \qquad
	Q_{01}^{\dagger} = \dot{y} \xi, 
	\qquad
	Z = A(F-\overline{F}).
\end{equation}
One may observe the vanishing $Z$ by using the equation of motion $ A = 0.$ 
On the other hand, $L_3$ has at least one dynamical variable of each $\Z2$-degree and the Noether charges after the use of equations of motion read as follows :
\begin{alignat}{3}
	Q_{10} &= \dot{y} \bar{\psi} + \dot{z} \bar{\xi}, & \qquad
	Q_{10}^{\dagger} &= \dot{y} \psi - \dot{\bar{z}} \xi,
	\nn \\
	Q_{01} &= \dot{y} \bar{\xi} + \dot{z} \bar{\psi}, & 
    Q_{01}^{\dagger} &= \dot{y} \xi - \dot{\bar{z}} \psi, & \qquad
    Z &= \dot{y} (\dot{z} + \dot{\bar{z}})
\end{alignat} 
which has non-vanishing $Z$. 
Finally, for $ L_4$ with no dynamical degree $(0,0)$ variables we have the conserved charges with vanishing $Z:$
\begin{equation}
	Q_{10} = \dot{z} \bar{\xi}, \qquad Q_{10}^{\dagger} = \dot{\bar{z}} \xi, \qquad
	Q_{01} = \dot{z} \bar{\psi}, \qquad Q_{01}^{\dagger} = \dot{\bar{z}} \psi, \qquad
	Z = 0. 
\end{equation}
This is a remarkable fact.  We started with the irreps of $\G$ having non-vanishing $Z$ and constructed classical actions invariant under the $\Z2$-SUSY transformations defined by the irreps. 
Nevertheless, we end up with the vanishing $Z$ if there is no dynamical variables of degree $(0,0)$ or $(1,1). $ 
This is not because of the theories discussed here are free. 
The vanishing $Z$ has also been observed in the $ {\cal N} = 1 $ models with an interaction term \cite{AKTqu}. 
Thus, this would be a universal feature of $\Z2$-graded mechanics.

%%%%%%%%%%%%%%%%%%%%%%%%%%%%%%%%%%%%%%%%%%%%%%%%%%%%%%%%%%%%%%%%%%%%%%%%%%%%%%%%%%%%
%
\setcounter{equation}{0}

\section{Concluding remarks} \label{SEC:CR}

%%%%%%%%%%%%%%%%%%%%%%%%%%%%%%%%%%%%%%%%%%%%%%%%%%%%%%%%%%%%%%%%%%%%%%%%%%%%%%%%%%%%
%
\setcounter{equation}{0}

We have investigated irreps of the $\Z2$-SUSY algebra of ${\cal N} = 2$ and found that irreps are eight dimensional if $ \lambda \neq E^2, $ while they are four dimensional if $ \lambda = E^2$ or $ \lambda = 0. $ 
The condition $ \lambda = E^2 $  also hold for all the $\Z2$-graded versions of quantum and classical mechanics discussed so far in the literature.  
We thus employed the four dimensional irreps ($\lambda = E^2$) as the $ \Z2$-SUSY transformation of $ {\cal N} = 2$ and construct several classical actions invariant under the transformation. 
It then turned out that even simple free theories are able to exhibit a distinct feature of $\Z2$-graded classical mechanics, i.e., vanishing the degree $(1,1)$ Noether charge $Z$. 
The disappearance of $Z$ was also observed in the interacting theory of ${\cal N} =1.$
This would be problematic when the theory is quantized since the known models of $\Z2$-graded SUSY quantum mechanics have non-vanishing $Z$ operator. 
To reproduce  $ {\cal N} = 1 $ quantum mechanics with non-vanishing $Z$, two distinct representations of the quantized operators are combined \cite{AKTqu}. 
One may apply the same technique to $ {\cal N} = 2 $ theories. 
It is, however, natural to anticipate the existence of more natural way of quantization. 

 We focused mainly on free theories in the present work. It is, of course, important to introduce interactions, but it is not a trivial problem to find $\Z2$-supersymmetric interaction terms. 
A possible idea of seeking interacting theories is the use of the $\Z2$-graded version of superfield. 
There is an obstacle to extend superfields to $\Z2$-grading, that is, integration on $\Z2$-graded superspace. 
Not all the functions on the superspace is integrable.  
This fact constrains possible form of $\Z2$-supersymmetric actions. 
The ${\cal N} =1$ case  has already been considered in \cite{DoiAi2}. 
Consideration of ${\cal N} =2$ is one of the works should be done. 
We remark that integration on graded supermanifolds itself is still an open problem of mathematics (see \cite{Pz2nint}). 

We also focused on $ \lambda = E^2$ irreps. 
In this case, the physical meaning of $\lambda $ is apparent. 
However, this is not a necessary condition. 
We have the eight dimensional irreps for non-vanishing $ \lambda \neq E^2$ which are able to transform more objects than four dimensional ones. Furthermore, we have one more parameter which we are able to control. 
This implies that the $\Z2$-SUSY is richer than the standard one. 
At this point, no physical systems having such $\Z2$-SUSY have been found, but searching of such an exotic SUSY is an interesting and challenging problem.    

 The present work is a first step to  $\Z2$-graded supersymmetry with higher ${\cal N}$. 
Knowledge on irreps of $\Z2$-SUSY algebra with higher ${\cal N}$ is desirable  to investigate such symmetries in physics and mathematics. However, to our best knowledge, no works on classification of irreps of $\Z2$-SUSY algebra for $ {\cal N} \geq 2$ has been done. 
As we saw in \S \ref{SEC:DE} and \S \ref{SEC:DEl}, irreps are sensitive to the eigenvalues of the center and the Casimir and more involved than the standard SUSY algebra. 
This is also one of the works should be done.

\appendix
\section{Action of $\G$}

In this Appendix, we give an explicit expression of the action of $\G$ on the basis vectors of $D(E,\lambda)$. 
The action of $H$ is diagonal and its representation matrix is the 32 dimensional scalar matrix whose diagonal entries are $E.$ 

\medskip\noindent
Action of $Q_{10}$
\begin{alignat}{2}
 \begin{pmatrix}
 	v_1 \\ v_2 \\ v_3 \\ v_4 \\ v_5 \\ v_6 \\ v_7 \\ v_8
 \end{pmatrix} 
 &\mapsto
 \begin{pmatrix}
   \chi_1 \\  -E\chi_1 \\ -E\chi_1 + \chi_4 -i\chi_5 \\ E(\chi_4 -i \chi_5) + \frac{i}{2}\chi_7 \\ 0 \\ i\lambda\chi_2 +2E\chi_6-\chi_8 \\ \chi_7 \\ i \lambda \chi_1
 \end{pmatrix},
 &\quad
 \begin{pmatrix}
	u_1 \\ u_2 \\ u_3 \\ u_4 \\ u_5 \\ u_6 \\ u_7 \\ u_8
 \end{pmatrix} 
 &\mapsto
 \begin{pmatrix}
 	0 \\ 2E\sigma_2 - \sigma_4 +i\sigma_6 \\ \sigma_3 \\ i\sigma_5 \\ \sigma_5 \\ -E\sigma_5 \\ -i\lambda \sigma_1+\sigma_8-E\sigma_5 \\ E (-i\lambda \sigma_1+\sigma_8) +\frac{i}{2} \lambda\sigma_3
 \end{pmatrix},
\\[5pt]
 \begin{pmatrix}
	\chi_1 \\ \chi_2 \\ \chi_3 \\ \chi_4 \\ \chi_5 \\ \chi_6 \\ \chi_7 \\ \chi_8
 \end{pmatrix} 
 &\mapsto
 \begin{pmatrix}
 	0 \\ \frac{1}{2}(Ev_1+v_2) \\ E(v_2-v_3)+v_4-\frac{i}{2}v_7 \\[3pt] \frac{i}{2}v_5 \\[3pt] \frac{1}{2}v_5 \\ \frac{1}{2}(-i\lambda  v_1+v_8) \\ 0 \\-\frac{i}{2}\lambda (Ev_1-v_2)+E v_8 
 \end{pmatrix},
 &
 \begin{pmatrix}
	\sigma_1 \\ \sigma_2 \\ \sigma_3 \\ \sigma_4 \\ \sigma_5 \\ \sigma_6 \\ \sigma_7 \\ \sigma_8
 \end{pmatrix} 
 &\mapsto
 \begin{pmatrix}
 	\frac{1}{2} u_1 \\ \frac{1}{2}(u_4-iu_5) \\ 0 \\ E u_4 -\frac{i}{2}(Eu_5-u_6) \\ 0 \\ \frac{1}{2}(Eu_5+u_6) \\ -\frac{i}{2}\lambda u_3 + E(u_6-u_7)+u_8 \\ \frac{i}{2}\lambda u_1
 \end{pmatrix}.
\end{alignat}
Action of $Q_{10}^{\dagger}$
\begin{alignat}{2}
	\begin{pmatrix}
		v_1 \\ v_2 \\ v_3 \\ v_4 \\ v_5 \\ v_6 \\ v_7 \\ v_8
	\end{pmatrix} 
	&\mapsto
	\begin{pmatrix}
       \chi_2 \\ E\chi_2 \\ E \chi_2 - \chi_3+i\chi_6 \\ \frac{i}{2} \chi_8 \\ i\lambda\chi_1+2E\chi_5-\chi_7 \\ 0 \\ i\lambda\chi_2 \\ \chi_8 
	\end{pmatrix},
	&\quad
	\begin{pmatrix}
		u_1 \\ u_2 \\ u_3 \\ u_4 \\ u_5 \\ u_6 \\ u_7 \\ u_8
	\end{pmatrix} 
	&\mapsto
	\begin{pmatrix}
       2E\sigma_1-\sigma_3+i\sigma_5 \\ 0 \\ i\sigma_6 \\ \sigma_4 \\ \sigma_6 \\ E\sigma_6 \\ i\lambda \sigma_2 +E\sigma_6-\sigma_7 \\ \frac{i}{2}\lambda\sigma_4 
	\end{pmatrix},
	\\[5pt]
	\begin{pmatrix}
		\chi_1 \\ \chi_2 \\ \chi_3 \\ \chi_4 \\ \chi_5 \\ \chi_6 \\ \chi_7 \\ \chi_8
	\end{pmatrix} 
	&\mapsto
	\begin{pmatrix}
		\frac{1}{2}(Ev_1-v_2) \\ 0 \\ \frac{i}{2}v_6 \\ v_4-\frac{i}{2}v_8 \\ \frac{1}{2} (-i\lambda v_1+v_7) \\ \frac{1}{2} v_6 \\ -\frac{i}{2}\lambda(Ev_1+v_2) + E v_7 \\ 0 
	\end{pmatrix},
	&
	\begin{pmatrix}
		\sigma_1 \\ \sigma_2 \\ \sigma_3 \\ \sigma_4 \\ \sigma_5 \\ \sigma_6 \\ \sigma_7 \\ \sigma_8
	\end{pmatrix} 
	&\mapsto
	\begin{pmatrix}
		\frac{1}{2}(u_3-iu_5) \\ \frac{1}{2}u_2 \\ Eu_3 -\frac{i}{2}(Eu_5+u_6) \\ 0 \\ \frac{1}{2}(Eu_5-u_6) \\ 0 \\ \frac{i}{2} \lambda u_2 \\ u_8 -\frac{i}{2} \lambda u_4
	\end{pmatrix}.
\end{alignat}
Action of $Q_{01}$
\begin{alignat}{2}
	\begin{pmatrix}
		v_1 \\ v_2 \\ v_3 \\ v_4 \\ v_5 \\ v_6 \\ v_7 \\ v_8
	\end{pmatrix} 
	&\mapsto
	\begin{pmatrix}
      \sigma_1 \\ -E\sigma_1+\sigma_3+i\sigma_5 \\ -E \sigma_1 \\ -\frac{i}{2}\sigma_8 \\ 0 \\ -i\lambda\sigma_2+2E\sigma_6-\sigma_7 \\ -i\lambda\sigma_1 \\ \sigma_8 
	\end{pmatrix},
	&\quad
	\begin{pmatrix}
		u_1 \\ u_2 \\ u_3 \\ u_4 \\ u_5 \\ u_6 \\ u_7 \\ u_8
	\end{pmatrix} 
	&\mapsto
	\begin{pmatrix}
      0 \\ 2E\chi_2-\chi_3-i\chi_6 \\ -i\chi_5 \\ \chi_4 \\ \chi_5 \\ i\lambda \chi_1-E\chi_5+\chi_7 \\ -E\chi_5 \\ -\frac{i}{2} \lambda \chi_4 
	\end{pmatrix},
	\\[5pt]
	\begin{pmatrix}
		\chi_1 \\ \chi_2 \\ \chi_3 \\ \chi_4 \\ \chi_5 \\ \chi_6 \\ \chi_7 \\ \chi_8
	\end{pmatrix} 
	&\mapsto
	\begin{pmatrix}
		\frac{1}{2}u_1 \\ \frac{1}{2}(u_3+iu_5) \\ Eu_3 + \frac{i}{2}(Eu_5-u_7) \\ 0 \\ 0 \\ \frac{1}{2}(Eu_5+u_7) \\ -\frac{i}{2} \lambda u_1 \\ \frac{i}{2}\lambda u_4+u_8
	\end{pmatrix},
	&
	\begin{pmatrix}
		\sigma_1 \\ \sigma_2 \\ \sigma_3 \\ \sigma_4 \\ \sigma_5 \\ \sigma_6 \\ \sigma_7 \\ \sigma_8
	\end{pmatrix} 
	&\mapsto
	\begin{pmatrix}
		0 \\ \frac{1}{2}(Ev_1+v_3) \\ -\frac{i}{2} v_5 \\ v_4+\frac{i}{2}v_8 \\ \frac{1}{2}v_5 \\ \frac{1}{2}(i\lambda v_1+v_7) \\ Ev_7 + \frac{i}{2} \lambda (Ev_1-v_3) \\ 0
	\end{pmatrix}.
\end{alignat}
Action of $Q_{01}^{\dagger}$
\begin{alignat}{2}
	\begin{pmatrix}
		v_1 \\ v_2 \\ v_3 \\ v_4 \\ v_5 \\ v_6 \\ v_7 \\ v_8
	\end{pmatrix} 
	&\mapsto
	\begin{pmatrix}
      \sigma_2 \\ E\sigma_2-\sigma_4-i\sigma_6 \\ E\sigma_2 \\ E(\sigma_4+i\sigma_6) -\frac{i}{2}\sigma_7 \\ -i\lambda\sigma_1+2E\sigma_5-\sigma_8 \\ 0 \\ \sigma_7 \\ -i\lambda\sigma_2 
	\end{pmatrix},
	&\quad
	\begin{pmatrix}
		u_1 \\ u_2 \\ u_3 \\ u_4 \\ u_5 \\ u_6 \\ u_7 \\ u_8
	\end{pmatrix} 
	&\mapsto
	\begin{pmatrix}
      2E\chi_1-\chi_4-i\chi_5 \\ 0 \\ \chi_3 \\ -i\chi_6 \\ \chi_6 \\ -i\lambda\chi_2+E\chi_6-\chi_8 \\ E\chi_6 \\ i\lambda(E\chi_2-\frac{1}{2}\chi_3)+E\chi_8 
	\end{pmatrix},
	\\[5pt]
	\begin{pmatrix}
		\chi_1 \\ \chi_2 \\ \chi_3 \\ \chi_4 \\ \chi_5 \\ \chi_6 \\ \chi_7 \\ \chi_8
	\end{pmatrix} 
	&\mapsto
	\begin{pmatrix}
		 \frac{1}{2}(u_4+iu_5) \\ \frac{1}{2}u_2 \\ 0 \\ Eu_4 + \frac{i}{2}(Eu_5+u_7) \\ \frac{1}{2}(Eu_5-u_7) \\ 0 \\ \frac{i}{2}\lambda u_3 + E(u_6-u_7)+u_8 \\ -\frac{i}{2} \lambda u_2
	\end{pmatrix},
	&
	\begin{pmatrix}
		\sigma_1 \\ \sigma_2 \\ \sigma_3 \\ \sigma_4 \\ \sigma_5 \\ \sigma_6 \\ \sigma_7 \\ \sigma_8
	\end{pmatrix} 
	&\mapsto
	\begin{pmatrix}
		\frac{1}{2}(Ev_1-v_3) \\ 0 \\ E(v_2-v_3) + v_4 + \frac{i}{2}v_7 \\ -\frac{i}{2}v_6 \\ \frac{1}{2}(i\lambda v_1+v_8)  \\ \frac{1}{2}v_6 \\ 0 \\ \frac{i}{2} \lambda(Ev_1+v_3) + Ev_8 
	\end{pmatrix}.
\end{alignat}
Action of $Z$
\begin{alignat}{2}
	\begin{pmatrix}
		v_1 \\ v_2 \\ v_3 \\ v_4 \\ v_5 \\ v_6 \\ v_7 \\ v_8
	\end{pmatrix} 
	&\mapsto
	\begin{pmatrix}
        u_5 \\ u_6 \\ u_7 \\ u_8 \\ \lambda u_1 \\ \lambda u_2 \\ \lambda u_3 \\ \lambda u_4
 	\end{pmatrix},
	&\qquad\qquad 
	\begin{pmatrix}
		u_1 \\ u_2 \\ u_3 \\ u_4 \\ u_5 \\ u_6 \\ u_7 \\ u_8
	\end{pmatrix} 
	&\mapsto
	\begin{pmatrix}
       v_5 \\ v_6  \\ v_7  \\ v_8 \\ \lambda v_1 \\ \lambda v_2\\ \lambda v_3 \\ \lambda v_4
	\end{pmatrix},
	\\[5pt]
	\begin{pmatrix}
		\chi_1 \\ \chi_2 \\ \chi_3 \\ \chi_4 \\ \chi_5 \\ \chi_6 \\ \chi_7 \\ \chi_8
	\end{pmatrix} 
	&\mapsto
	\begin{pmatrix}
		-\sigma_5 \\ -\sigma_6   \\ -\sigma_7 \\ -\sigma_8  \\ -\lambda \sigma_1  \\ -\lambda \sigma_2  \\ -\lambda \sigma_3  \\ -\lambda \sigma_4
	\end{pmatrix},
	&
	\begin{pmatrix}
		\sigma_1 \\ \sigma_2 \\ \sigma_3 \\ \sigma_4 \\ \sigma_5 \\ \sigma_6 \\ \sigma_7 \\ \sigma_8
	\end{pmatrix} 
	&\mapsto
	\begin{pmatrix}
		-\chi_5 \\ -\chi_6  \\ -\chi_7 \\ -\chi_8 \\ -\lambda\chi_1  \\ -\lambda\chi_2 \\ -\lambda\chi_3 \\ -\lambda\chi_4
	\end{pmatrix}.
\end{alignat}

One may read off the action of $ \G$ on the basis vector of $ D(E)$ by setting $ \lambda = 0 $ and restricting to the vectors $ v_i, u_i, \chi_i, \sigma_i $ with $i = 1, 2, 3, 4.$ 
Therefore, $Z$ is represented by the zero matrix in $D(E). $ 

%%%%%%%%%%%%%%%%%%%%%%%%%%%%%%%%%%%%%%%%%%%%%%%%%%%%%%%%%%%%%%%%%%%%%%%%%%%%%

\section*{Author's contributions}

All authors contributed equally to this work.

\section*{Data availability}

Data sharing is not applicable to this article as no new data were created or analyzed in this study.

%%%%%%%%%%%%%%%%%%%%%%%%%%%%%%%%%%%%%%%%%%%%%%%%%%%%%%%%%%%%%%%%%%%%%%%%%%%%%
%
%   References
%


\begin{thebibliography}{99}
\bibitem{Ree} R. Ree, 
\textit{Generalized Lie elements}, Canad. J. Math. \textbf{12}, 493 (1960).

\bibitem{sch1} M. Scheunert, \textit{Generalized Lie algebras}, J. Math. Phys. {\bf 20}, 712 (1979).

\bibitem{RW1} V. Rittenberg and D. Wyler, {\it  Generalized superalgebras}, Nucl. Phys. B {\bf 139}, 189 (1978).

\bibitem{RW2} V. Rittenberg and D. Wyler, {\it Sequences of $\mathbb{Z}_2 \otimes \mathbb{Z}_2$ graded Lie algebras and superalgebras}, 
J. Math. Phys. {\bf 19}, 2193  (1978).
	
%%%%%%%%%%%%%%%%%%%%%%%%%%%%%%%%%%%%%%%%%%%%%%%%%%%%%%%%%%%%%%
\bibitem{Vasiliev} 
M. A. Vasiliev,
\textit{de Sitter supergravity with positive cosmological constant and generalised Lie superalgebras}, 
 Class. Quantum Grav. \textbf{2}, 645 (1985).	

\bibitem{tol} V. N. Tolstoy, 
\textit{Once more on parastatistics}, Phys. Part. Nucl. Lett. {\bf{11}},  933 (2014).


\bibitem{AKTT1} N. Aizawa,  Z. Kuznetsova, H. Tanaka, F. Toppan,
\textit{${\mathbb Z}_2\times {\mathbb Z}_2$-graded Lie Symmetries of the L\'evy-Leblond Equations}, 
Prog. Theor. Exp. Phys. \textbf{2016}, 123A01 (2016).

\bibitem{AKTT2} N. Aizawa,  Z. Kuznetsova, H. Tanaka, F. Toppan,
\textit{Generalized supersymmetry and L\'evy-Leblond equation}, 
in ``Physical and Mathematical Aspects of Symmetries,''
J.-P. Gazeau, S. Faci, T. Micklitz, R. Scherer, F. Toppan (editors),  
Springer (2017) p.79.

\bibitem{tol2} V. N. Tolstoy,  
\textit{Super-de Sitter and Alternative Super-Poincar\'e Symmetries},  
In: Dobrev V. (eds) Lie Theory and Its Applications in Physics. Springer Proceedings in Mathematics \& Statistics, 
vol. 111, Springer, Tokyo, 2014.

	
\bibitem{Bruce} A. J. Bruce, 
\textit{On a $\mathbb{Z}_2^n$-graded version of supersymmetry}, Symmetry \textbf{11}, 116 (2019).

\bibitem{BruDup} A. J. Bruce and S. Duplij, 
{\it Double-graded supersymmetric quantum mechanics},
J. Math. Phys.  \textbf{61}, 063503 (2020).

\bibitem{AAD}N. Aizawa, K. Amakawa, S. Doi, 
{\it $\mathcal{N}$-Extension of double-graded supersymmetric and superconformal quantum mechanics}, 
J. Phys. A: Math. Theor. \textbf{53}, 065205 (2020).


\bibitem{AAd2} N. Aizawa, K. Amakawa, S. Doi, 
{\it $\mathbb{Z}_2^n$-Graded extensions of supersymmetric quantum mechanics via Clifford algebras}, 
J. Math. Phys. \textbf{61}, 052105 (2020).


\bibitem{AKTcl} N. Aizawa, Z. Kuznetsova and F. Toppan,
{\it ${\mathbb Z}_2\times {\mathbb Z}_2$-graded mechanics: the classical theory},  
Eur. Phys. J. C \textbf{80}, 668 (2020).

\bibitem{AKTqu} N. Aizawa, Z. Kuznetsova and F. Toppan,
{\it ${\mathbb Z}_2\times {\mathbb Z}_2$-graded mechanics: the quantization},  
Nucl. Phys. B \textbf{967},  115426 (2021).


\bibitem{DoiAi1} S. Doi and N. Aizawa,
\textit{$\mathbb{Z}_2^3$-Graded extensions of Lie superalgebras and superconformal quantum mechanics}, 
SIGMA	\textbf{17} 071, (2021).

\bibitem{DoiAi2} S. Doi and N. Aizawa,
\textit{Comments of $\mathbb{Z}_2^2$-supersymmetry in superfield formalism}, 
Nucl. Phys.  \textbf{B974}, 115641 (2022).

\bibitem{brusigma} A. J. Bruce, \textit{$\mathbb{Z}_2 \times \mathbb{Z}_2$-graded supersymmetry: 2-d sigma models},
J. Phys. A:Math. Theor. \textbf{53}, 455201 (2020).

\bibitem{bruSG} A. J. Bruce, \textit{Is the $\mathbb{Z}_2 \times \mathbb{Z}_2$-graded sine-Gordon equation integrable ?}, 
Nucl. Phys. B \textbf{971}, 115514 (2021).


\bibitem{Topp} F. Toppan, {\it $\mathbb{Z}_{2} \times \mathbb{Z}_{2}$-graded parastatics in multiparticle quantum Hamiltonians}, J. Phys. A: Math. Theor. \textbf{54}, 115203 (2021).
%arXiv:2008.11554 [hep-th].

\bibitem{Topp2} F. Toppan, \textit{Inequivalent quantizations from gradings and $ \mathbb{Z}_2 \times \mathbb{Z}_2$ parabosons}, 
J. Phys. A, Math. Theor. \textbf{54}, 355202 (2021).
%	arXiv:2104.09692 [hep-th].

  %%%%%%%%%%%%%%%%%%%%%%%%%%%%%%%%%%%%%%%%%%%%%%%%%

\bibitem{CGP1} T. Covolo, J. Grabowski and N. Poncin, 
%\textit{$ \mathbb{Z}_2^n$-Supergeometry I: Manifolds and Morphisms},  arXiv:1408.2755 [math.DG].
\textit{The category of $\mathbb{Z}_2^n$-supermanifolds}, 
J. Math. Phys. \textbf{57}, 073503 (2016).


\bibitem{CGP2} T. Covolo, J. Grabowski and N. Poncin, 
%\textit{$ \mathbb{Z}_2^n$-Supergeometry II: Batchelor-Gawedzki Theorem}, arXiv:1408.2939 [math.DG].
\textit{Splitting theorem for $ \mathbb{Z}_2^n$-supermanifolds}, 
J. Geom. Phys. \textbf{110}, 393 (2016).

\bibitem{CGP3} T. Covolo, V. Ovsienko, and N. Poncin, 
\textit{Higher trace and Berezinian of matrices over a Clifford algebra}, J. Geom. Phys. \textbf{62}, 2294 (2012).


\bibitem{CGP4} M. Mohammadi and S. Varsaie, 
\textit{On the construction of $\mathbb{Z}_2^n$-grassmannians as homogeneous $\mathbb{Z}_2^n$-spaces},  
Electron. Res. Arch.  \textbf{30}, 221 (2022). 

\bibitem{CoKPo} T. Covolo, S. Kwok and N. Poncin, 
\textit{Differential calculus on  $\mathbb{Z}_2^n$-supermanifolds}, arXiv:1608.00949 [math.DG].

\bibitem{Pz2nint} N. Poncin, \textit{Towards integration on colored supermanifolds},
Banach Cent. Publ. \textbf{110}, 201 (2016).


\bibitem{BruIbar} 
A. J. Bruce and E. Ibarguengoytia
\textit{The graded differential geometry of mixed symmetry tensors}, 
\textit{Arch. Math.} (Brno) \textbf{55}, 123 (2019). %; arXiv:1806.04048[math.ph].	


\bibitem{BruPon} A. J. Bruce and N. Poncin, 
\textit{Functional analytic issues in $ \mathbb{Z}_2^n$-geometry}, 
Rev. Un. Mat. Argentina \textbf{60}, 611 (2019).

\bibitem{BruIbarPonc} A. J. Bruce, E. Ibarguengoytia and N. Poncin, 
\textit{The Schwarz--Voronov Embedding of ${\mathbb Z}_{2}^{n}$-Manifolds}, 
SIGMA \textbf{16}, 002 (2020).

\bibitem{BruGraRiemann} A. J. Bruce and J. Grabowksi, 
\textit{Riemannian structures on  $\mathbb{Z}_2^n$-manifolds}, 
Mathematics \textbf{8}, 1469 (2020).


\bibitem{BruGrabow} A. J. Bruce and J. R. Grabowski, 
\textit{Odd connections on supermanifolds: Existence and relation with affine connections},
J. Phys. A:Math. Theor. \textbf{53}, 455203 (2020).

\bibitem{CoKwPon} T. Covolo, S. Kwok, N. Poncin,
\textit{Local forms of morphisms of colored supermanifolds}, 
J. Geom. Phys. \textbf{168}, 104302 (2021).

\bibitem{BruIbarPon2} A. J. Bruce, E. Ibarguengoytia and N. Poncin, 
\textit{Linear $\mathbb{Z}_2^n$-Manifolds and Linear Actions}, 
SIGMA \textbf{17}, 060 (2021).  

\bibitem{BruGrabow2} A. J. Bruce, J. Grabowski,  
\textit{Symplectic $\mathbb{Z}_2^n$-manifolds}, 
J. Geom. Mech. \textbf{13}, 285 (2021). 

%%%%%%%%%%%%%%%%%%%%%%%%%%%%%%%%%%%%%%%%%%%%%%%%%%%%%%%%%%

\bibitem{MohSal} M. Mohammadi and H.  Salmasian, \textit{The Gelfand-Naimark-Segal construction for unitary representations of $\mathbb Z_2^n$-graded Lie supergroups}, Banach Cent. Publ. \textbf{113}, 263 (2017).

   \bibitem{NAJap} N. Aizawa, 
\textit{Verma modules over a ${\mathbb Z}_2\times {\mathbb Z}_2$ graded superalgebra and invariant differential equations}, 
Scientiae Mathematicae Japonicae, \textbf{31}, 2018-4 (2018).   

\bibitem{IsStvdJ} P. S. Isaac, N. I. Stoilova, J. van der Jeugt, 
\textit{The $\mathbb{Z}_2 \times \mathbb{Z}_2$-graded general linear Lie superalgebra},
J. Math. Phys. \textbf{61}, 011702 (2020).

 \bibitem{Meyer} P. Meyer, 
\textit{The Kostant invariant and special $\epsilon$-orthogonal representations for $\epsilon$--quadratic colour Lie algebras}, 
J. Alg. \textbf{572}, 337 (2021).

\bibitem{AmaAi} K. Amakawa and N. Aizawa,
\textit{A classification of lowest weight irreducible modules over $\mathbb{Z}_2^2$-graded extension of $osp(1|2)$}, 
J. Math. Phys.  \textbf{62}, 043502 (2021). 


\bibitem{Que} C. Quesne, \textit{Minimal bosonization of double-graded supersymmetric quantum mechanics}, 
Mod. Phys. Lett. A \textbf{36},  2150238 (2021).

\bibitem{PT} A. Pashnev and F. Toppan, \textit{On the classification of $N$-extended supersymmetric quantum mechanics}, J. Maht.  Phys. \textbf{42}, 5257 (2001).


\bibitem{KRT} Z. Kuznetsova, M. Rojas and F. Toppan, 
\textit{Classification of irreps and invariants of the $N$-extended supersymmetric quantum mechanics}, JHEP \textbf{0603}, 098 (2006).  




\end{thebibliography}
\end{document}